%
%
%
\def\today{\ifcase\month\or January\or February\or March\or April\or May\or
June\or July\or August\or September\or October\or November\or December\fi
\space\number\day, \number\year}
%
%
\newcount\notenumber

\def\note{\global\advance\notenumber by 1 \footnote{$^{\the\notenumber}$}}
%
%
\newif\ifsectionnumbering
\newcount\eqnumber
\def\cleareqnumber{\eqnumber=0}
\def\numbereq{\global\advance\eqnumber by 1
\ifsectionnumbering\eqno(\the\secnumber.\the\eqnumber)\else\eqno
(\the\eqnumber)\fi}
\def\eqalinno{{\global\advance\eqnumber by 1}
\ifsectionnumbering(\the\secnumber.\the\eqnumber)\else(\the\eqnumber)\fi}
\def\name#1{\ifsectionnumbering\xdef#1{\the\secnumber.\the\eqnumber}
\else\xdef#1{\the\eqnumber}\fi}
\def\nosectionnumbering{\sectionnumberingfalse}
\sectionnumberingtrue
%
%
\newcount\refnumber

\immediate\openout1=refs.tex
\immediate\write1{\noexpand\frenchspacing}
\immediate\write1{\parskip=0pt}
\def\ref#1#2{\global\advance\refnumber by 1%
[\the\refnumber]\xdef#1{\the\refnumber}%
\immediate\write1{\noexpand\item{[#1]}#2}}
\def\tie{\noexpand~}

%
%
\font\twelvebf=cmbx10 scaled \magstep1
\newcount\secnumber

\def\newsection#1.{\ifsectionnumbering\cleareqnumber\else\fi%
	\global\advance\secnumber by 1%
	\bigbreak\bigskip\par%
	\line{\twelvebf \the\secnumber. #1.\hfil}\nobreak\medskip\par\noindent}
%
%
%
\def \sqr#1#2{{\vcenter{\vbox{\hrule height.#2pt
	\hbox{\vrule width.#2pt height#1pt \kern#1pt
		\vrule width.#2pt}
		\hrule height.#2pt}}}}

%
%
%
\newdimen\fullhsize
\def\fiddle{\fullhsize=6.5truein \hsize=3.2truein}
\def\fullline{\hbox to\fullhsize}
\def\mkhdline{\vbox to 0pt{\vskip-22.5pt
	\fullline{\vbox to8.5pt{}\the\headline}\vss}\nointerlineskip}
\def\mkftline{\baselineskip=24pt\fullline{\the\footline}}
\let\lr=L \newbox\leftcolumn
\def\twocolumns{\fiddle
	\output={\if L\lr \global\setbox\leftcolumn=\columnbox
		\global\let\lr=R \else \doubleformat \global\let\lr=L\fi
		\ifnum\outputpenalty>-20000 \else\dosupereject\fi}}
\def\doubleformat{\shipout\vbox{\mkhdline
		\fullline{\box\leftcolumn\hfil\columnbox}
		\mkftline} \advancepageno}
\def\columnbox{\leftline{\pagebody}}
\nosectionnumbering
\magnification=1200
\def\pr#1 {Phys. Rev. {\bf D#1\tie }}
\def\pe#1 {Phys. Rev. {\bf #1\tie}}
\def\pre#1 {Phys. Rep. {\bf #1\tie}}
\def\pl#1 {Phys. Lett. {\bf #1B\tie }}
\def\prl#1 {Phys. Rev. Lett. {\bf #1\tie }}
\def\np#1 {Nucl. Phys. {\bf B#1\tie }}
\def\ap#1 {Ann. Phys. (NY) {\bf #1\tie }}
\def\cmp#1 {Commun. Math. Phys. {\bf #1\tie }}
\def\imp#1 {Int. Jour. Mod. Phys. {\bf A#1\tie }}
\def\mpl#1 {Mod. Phys. Lett. {\bf A#1\tie}}
\def\jhep#1 {JHEP {\bf #1\tie}}
\def\nuo#1 {Nuovo Cimento {\bf B#1\tie}}
\def\tie{\noexpand~}

\parskip=15pt plus 4pt minus 3pt
\headline{\ifnum \pageno>1\it\hfil Possible extensions of the
4-D Schwarzschild horizon $\ldots$\else \hfil\fi}
\font\title=cmbx10 scaled\magstep1
\font\tit=cmti10 scaled\magstep1
\footline{\ifnum \pageno>1 \hfil \folio \hfil \else
\hfil\fi}
\raggedbottom


\overfullrule0pt


\rightline{\vbox{\hbox{RU00-09-B}\hbox{hep-th/0010183}}}
\vfill
\centerline{\title POSSIBLE EXTENSIONS OF THE 4-D SCHWARZSCHILD HORIZON}
\centerline{\title IN THE 5-D BRANE WORLD}
\vfill
{\centerline{\title Ioannis Giannakis${}^{a}$
and Hai-cang Ren${}^{a, b}$ \footnote{$^{\dag}$}
{\rm e-mail: \vtop{\baselineskip12pt
\hbox{giannak@theory.rockefeller.edu, ren@theory.rockefeller.edu,}}}}
}
\medskip
\centerline{$^{(a)}${\tit Physics Department, The Rockefeller
University}}
\centerline{\tit 1230 York Avenue, New York, NY
10021-6399}
\medskip
\centerline{$^{(b)}${\tit Department of Natural Science, Baruch College of CUNY}}
\centerline{\tit New York, NY 10010} 
\vfill
\centerline{\title Abstract}
\bigskip
{\narrower\narrower
We show that, in the absence of matter in the bulk, the Einstein
equations and the Gauss-normal form
of the metric place stringent restrictions on
the form of the event horizon in a brane world. As a
consequence, the off-brane extension of the standard
4-D Schwarzschild horizon in the Randall-Sundrum $AdS_5$
spacetime,
as it is viewed from the brane can only be of
a tubular shape,
instead of a pancake shape. When it is viewed from
the $AdS_5$ horizon, such a tubular horizon is absent. 
\par}
\vfill\vfill\break



A recent breakthrough in high energy physics is the
possibility of observing large extra space dimensions.
It has been conjectured that the Standard Model particles
are confined to propagate within a 3-D brane embedded in a space
of $(4+d)$-D dimensions. On the contrary gravitons
can escape and propagate also in the bulk. The 
most important recognition is that the energy scale of the
extra dimensions can be as low as few TeV and their signature
may show up in future colliders (see for
example \ref\pesk{S. Cullen, M. Perelstein
and M. Peskin, \pr62 (2000) 055012.}). Since gravity is
formulated by $(4+d)$-D General Relativity,
the gravitational force within the brane may contain
detectable corrections to Newton's law
\ref{\price}{J. C. Price, in proc. Int. Symp. on
Experimental Gravitational Physics, ed. P. F. Guangzhou, China
( World Scientific, Singapore 1988 ); J. C. Price et. al.,
NSF proposal 1996; A. Kapitulnik and T. Kenny, NSF proposal, 1997.}.
The recovery of General Relativity on the {\it physical brane}
has become an interesting issue to explore.

One implementation of such a brane world scenario was
proposed by Randall and Sundrum \ref{\randall}{
L. Randall and R. Sundrum, \prl83 (1999) 4690.}.
The {\it physical brane} in their model is the junction
of two pieces of 5-D spacetime manifolds that
are asymptotically {\it anti-de Sitter}. The Gauss-normal form
of the metric in this space is
$$
ds^2=e^{-2\kappa|y|}{\bar g}_{\mu\nu}dx^\mu dx^\nu+dy^2,
\numbereq\name{\eqena}
$$
with the brane located at $y=0$ and $\kappa > 0$ sets
the energy scale of the extra space dimension. The metric
${\bar g}_{\mu\nu}$ is determined by the 5-D Einstein equations
$$
R_{mn}-{1\over 2}Rg_{mn}-\Lambda g_{mn}=-4{\pi}^2G_5
T_{\mu\nu}\delta^{\mu}_{m}\delta^{\nu}_{n}{\delta}(y)
\numbereq\name{\eqdyo}
$$
where the cosmological constant $\Lambda=-6\kappa^2$,
$G_5$ is the 5-D gravitational constant and $T_{\mu\nu}$
the energy-momentum tensor on the brane. Here and
throughout the 
paper, we adopt the convention that the Greek indices take values 
0-3 and the Latin indices 0-4. The 4-D gravitational
constant is given by $G \sim {{G_5}\over {\kappa}}$.
In the absence of matter, $T_{\mu\nu}=0$, ${\bar g}_{\mu\nu}=
\eta_{\mu\nu}$ is a solution of equation (\eqdyo).
Subsequently, the metric in equation (\eqena) becomes
that of $AdS_5$. The solution to equation (\eqdyo) can
also be obtained from the solution to the sourceless equation,
$T_{\mu\nu}=0$, subject to the appropriate Israel matching condition
\ref{\isr}{W. Israel, \nuo44
(1966) 1.} determined by $T_{\mu\nu}$.
The perturbative solution of (\eqdyo) to the linear
order in $G$ for an arbitrary $T_{\mu\nu}$ 
\ref{\giddings}{S. Giddings, E. Katz
and L. Randall, \jhep0003 (2000) 023.}
and to second
order $G^2$ for a static spherical mass distribution on
the brane \ref{\ren}{I. Giannakis and H. C. Ren, {\it
Recovery of the Schwarzschild metric in theories with localized
gravity beyond linear order}, hep-th/0007053,
to appear in Phys. Rev. D.}
reveals no tangible disagreement with the
classical tests of 4-D General Relativity at large $\kappa$.

It is generally believed that
4-D general relativity is recovered beyond the weak 
coupling expansion on the {\it physical brane} for large $\kappa$
\ref{\maeda}{T. Shiromizu, K. Maeda and M. Sasaki, \pr62
(2000) 024012.}.
By combining the Schwarzschild
metric on the {\it physical brane}
and the profile of the linear gravitational potential off the
{\it physical brane}, the authors of [\giddings] conjectured
a pancake shaped horizon for the physical
black hole, a gravitational field
generated by a mass point on the brane.
In this letter, we shall make some
rigorous statements on the form of the
horizon of a physical black hole, which when confined to the
{\it physical brane} reproduces the standard 4-D Schwarzschild
horizon. We find that the Einstein equations together with
a Gauss-normal form of the metric imply particular types of horizons
but the pancake shape does not belong to these types.

The  5-D Einstein sourceless equations can be rewritten as
$$
R_{\mu\nu}-4\kappa^2g_{\mu\nu}=0, \quad R_{y\mu}=0,
\quad R_{yy}-4\kappa^2=0,
\numbereq\name{\eqtesse}
$$
The most general metric in $D=4+1$ dimensions produced by a static,
spherically symmetric matter distribution on the {\it physical brane} 
has the following form:
$$
ds^2=e^{-2\kappa|y|}(-e^adt^2+e^bdr^2+e^cr^2d\Omega^2)+dy^2,
\numbereq\name{\eqpente}
$$
where $d\Omega^2=d\theta^2+\sin^2\theta d\phi^2$ is the solid angle on 
$S^2$ and $a$, $b$ and $c$ are functions of $r$ and $y$.
Substituting the metric (\eqpente) into
equations (\eqtesse), we obtain the following
components of the Einstein equation outside the source:
$$
\eqalign{
R_{tt}+4\kappa^2 e^{-2\kappa y+a}&=
{1\over 2}e^{a-b}\Big[-a^{\prime\prime}-{2\over r}a^\prime
+{1\over 2}a^\prime
(-a^\prime+b^\prime-2c^\prime)\cr
&-{1\over 2}e^{-2\kappa y+a}\Big[\ddot a-
5\kappa\dot a-\kappa\dot b-2\kappa
\dot c +{1\over 2}\dot a(\dot a+\dot b+2\dot c)\Big]=0\cr}
\numbereq\name{\eqevon}
$$
$$
\eqalign{
R_{rr}-4\kappa^2e^b={1\over 2}a^{\prime\prime}
&+c^{\prime\prime}-{1\over r}b^\prime+{2\over r}c^\prime
+{1\over 4}a^\prime(a^\prime-b^\prime)
-{1\over 2}c^\prime(b^\prime-c^\prime)\cr
&+{1\over 2}e^{-2\kappa y+b}\Big[\ddot b-5\kappa\dot b-\kappa\dot a
-2\kappa\dot c+{1\over 2}\dot b(\dot a+\dot b+2\dot c)\Big]=0\cr}
\numbereq\name{\eqasho}
$$
$$
\eqalign{
R_{\theta\theta}&-4\kappa^2 r^2e^c={1\over 2}r^2e^{c-b}\Big[
c^{\prime\prime}+{4\over r}c^\prime
+{{a^\prime-b^\prime}\over r}+c^{\prime2}
+{1\over 2}(a^\prime-b^\prime)c^\prime\Big]\cr
&+{1\over 2}r^2e^{-2\kappa y+c}\Big[
\ddot c-\kappa(\dot a+\dot b)-6\kappa \dot c
+{1\over 2}\dot c(\dot a+\dot b
+2\dot c)\Big]+e^{c-b}-1=0\cr}
\numbereq\name{\eqhbcv}
$$
$$
\eqalign{
&R_{yy}-4\kappa^2={1\over 2}(\ddot a+\ddot b+2\ddot c)
-\kappa(\dot a+\dot b+2\dot c)+{1\over 4}(\dot a^2
+\dot b^2+2\dot c^2)=0\cr
&R_{ry}=R_{yr}={1\over 2}\Big[\dot a^\prime
+2\dot c^\prime-{2\over r}(\dot b-\dot c)+{1\over 2}a^\prime
(\dot a-\dot b)-c^\prime(\dot b-\dot c)\Big]=0,\cr}
\numbereq\name{\eqefta}
$$
where the prime denotes a partial derivative with respect to $r$ and the 
dot denotes a partial derivative with respect to $y$. These equations apply 
to the positive side of the brane, $y>0$, the corresponding
equations to the negative side of the brane, $y<0$, are obtained
by switching the sign of $\kappa$.

Let us formulate the concept that would define an event horizon in the
5-D brane world. We assume that there 
exists a solution of the 5-D Einstein equations
which satisfies the
Israel matching condition across the brane and maintains the Lorentzian
signature in a certain region, ${\cal P}$,
of the parametric $r-y$ plane. We call ${\cal P}$ the physical region. 
The physical region may or may not cover the entire $r-y$ plane. 
An example of a physical region
that does not cover the entire parametric 
space is the $AdS$ $C$-metric discussed in
\ref{\myers}{R. Emparan,
G. Horowitz and R. Myers, \jhep0001 (2000) 007.}.
We also
assume that there exists an asymptotic region
${\cal A}$ within ${\cal P}$, where the metric components $e^a$, 
$e^b$ and $e^c$ are well approximated by linear gravity and the functions 
$a$, $b$ and $c$ are well behaved beyond ${\cal A}$. As we have seen, 
this is the case for both brane-based coordinates and for
$AdS_5$ horizon-
based coordinates. Starting from ${\cal A}$, we trace all possible light 
rays given by $ds^2=0$ or more specifically by
$$
dt^2=e^{b-a}dr^2+e^{c-a}r^2d\Omega^2+e^{2\kappa y-a}dy^2
\numbereq\name{\eqaristeros}
$$
until we come to a point from which the light can not propagate forward in 
certain spatial direction. By continuity, the union of these points forms 
a 4-D hypersurface, ${\cal H}$, which we refer to as an event horizon. We 
shall consider the case that ${\cal H}$
lies within (not coincide with the border of)
${\cal P}$.

To prevent light propagation, some of the coefficients on the right hand 
side of (\eqaristeros) need to become sufficiently
divergent so that an increment of the 
corresponding spatial coordinate would take infinite amount of time, $t$. 
Therefore, we can determine the form of the horizon
by finding the locus of the logarithmic
singularities of the functions $a$, $b$ and $c$. This locus has to be 
consistent with the Einstein equations (\eqevon)-(\eqefta).

We denote the locus of the logarithmic singularities of $a$, $b$ and $c$ 
by $H(r, y)=0$ and an arbitrary 
point on it by $P(r_0, y_0)$. The unit normal vector $\vec n$ and the unit 
tangent vector $\vec t$ at $P$ on $r-y$ plane are
$$
\vec n = (\cos\alpha, \sin\alpha), \qquad \vec t = (-\sin\alpha, \cos\alpha),
\numbereq\name{\eqkenderis}
$$
where 
$$
\cos\alpha = {1\over {\Delta}}\Big({{\partial H}\over {\partial r}}\Big)_P,
\qquad \sin\alpha ={1\over {\Delta}}
\Big({{\partial H}\over {\partial y}}\Big)_P
\numbereq\name{\eqthanou}
$$
and 
$$
\Delta = \sqrt{\Big({{\partial H}\over {\partial r}}\Big)_P^2
+\Big({{\partial H}\over {\partial y}}\Big)_P^2}.
\numbereq\name{\eqxantop}
$$
Let's consider a point $Q(r, y)$ in the neighbourhood of $P$ and
let's transform the 
coordinates into 
$$
\eqalign{
\xi&=(r-r_0)\cos\alpha+(y-y_0)\sin\alpha\cr
\eta&=-(r-r_0)\sin\alpha+(y-y_0)\cos\alpha.\cr}
\numbereq\name{\eqkarnat}
$$
As $\xi\to 0$ and $\eta\to 0$, we expect that
$$
e^a\simeq A\xi^{n_a}, \qquad e^b\simeq B{\xi}^{n_b}, 
\qquad e^c\simeq C{\xi}^{n_c}
\numbereq\name{\eqavios}
$$
where $n_a$, $n_b$ and $n_c$ are integers 
due to the reality requirement of the metric components on
both sides of the horizon, and $A$, $B$ and $C$
are numerical constants.
The integers
$n_a$, $n_b$ and $n_c$ could not all be zero,
otherwise there would be no singularity at the point $P$. 
The Lorentzian signature in the physical region forbids
$e^c$ and the metric determinant $e^{-8\kappa y+a+b+2c}r^4\sin^2\theta$
from changing their signs. These considerations
restrict both $n_c$ and $n_a+n_b+2n_c$ to be even. By continuity, these 
exponents are maintained along the entire $H$. Therefore,
$$
a\simeq n_a\ln\xi, \qquad b\simeq n_b\ln\xi, \qquad c\simeq n_c\ln\xi.
\numbereq\name{\eqverouli}
$$
Taking into account that
$$
\eqalign{
{{\partial}\over {{\partial r}}}&=\cos\alpha{{\partial}
\over {{\partial{\xi}}}}
-\sin\alpha{{\partial}\over {{\partial{\eta}}}}\cr
{{\partial}\over {{\partial y}}}&=\sin\alpha{{\partial}
\over {{\partial{\xi}}}}
+\cos\alpha{{\partial}\over {{\partial{\eta}}}},\cr}
\numbereq\name{\eqkoffa}
$$
their derivatives behave as
$$
\dot a\simeq {n_a\over {\xi}}\sin\alpha, \qquad \dot b\simeq
{n_b\over{\xi}}\sin\alpha, \qquad \dot c\simeq{n_c\over\xi}\sin\alpha
\numbereq\name{\eqpyrros}
$$
and
$$
a^\prime\simeq {n_a\over {\xi}}\cos\alpha, \qquad b^\prime\simeq
{n_b\over{\xi}}\cos\alpha, \qquad c^\prime\simeq{n_c\over\xi}\cos\alpha.
\numbereq\name{\eqpyrrote}
$$

We observe that the derivatives of $a, b, c$ become singular as $Q$
approaches $P$. As a result
of substituting (\eqpyrros) and (\eqpyrrote)
into equations (\eqevon)-(\eqefta), 
the cancellation of the leading sigularities in $R_{yy}-4\kappa^2$ and 
in $R_{ry}$, $O({1\over {\xi^2}})$,
leads to two conditions on $n_a$, $n_b$, $n_c$ and $\alpha$:
$$
E\sin^2\alpha=0, \qquad F\sin\alpha\cos\alpha=0,
\numbereq\name{\eqrigas}
$$
with
$$
E=-n_a-n_b-2n_c+{1\over 2}(n_a^2+n_b^2+2n_c^2)
\numbereq\name{\eqrhgaios}
$$
and 
$$
F=-n_a-2n_c+{1\over 2}n_a(n_a-n_b)-n_c(n_b-n_c).
\numbereq\name{\eqalhths}
$$
There are three types of solutions to be analyzed.

\noindent
{\it{Type I}}: $E\neq 0$.

The only solution is $\sin\alpha = 0$, which implies a tube shaped $H$: 
$$
\Big({\partial H\over \partial y}\Big)_P = 0.
\numbereq\name{\eqlymber}
$$
The standard Schwarszchild horizon of a physical
black hole belongs to this 
type. In the coordinates straight with respect
to the brane, the recovery of
4-D general relativity on the brane for $\kappa GM>>1$ 
implies that
$$
ds^2\simeq -(1-{2GM\over r})dt^2+{dr^2\over 1-{2GM\over r}}+r^2 d\Omega^2
\numbereq\name{\eqassions}
$$
for $y=0$. The standard Schwarzschild horizon at $r\simeq 2GM$
corresponds to the exponents $n_a=1$,
$n_b=-1$ and $n_c=0$.
Therefore, since
the integer combination $E\neq 0$, the off-brane extention 
of the horizon is a tube as is shown in Fig. 1a. It
either extends to infinity in the $y$ direction, 
similar to the horizon of the black cigar solution of
Chamblin-Hawking-Reall \ref{\haw}{A. Chamblin, S. W. Hawking
and H. S. Reall, \pr61 (2000) 065007.},
(though the explicit
forms of the solution are different), 
or it terminates at the border of the physical region,
similar to the example
in [\myers].   
This rules out the possibility of a pancake shaped 
Schwarzschild horizon in 5-D.
On the other hand, in the coordinates straight with respect to the $AdS_5$
horizon, the validity of linear gravity for large $r$ or large positive 
$y$ excludes the tubular form of the horizon completely, as is shown in 
Fig. 1b. Because of the brane 
bending in the negative $y$-direction
and the failure of the linear approximation there, 
we are unable to rule out the possibility of horizons of the
subsequent two
types in this coordinate system.

Before analyzing the next two types, we notice that the solutions to $E=0$
correspond to points with integer coordinates lying on an oblate spheroid 
with axis 2 and $\sqrt{2}$. There are only twelve of them, and we should 
exclude the solutions with odd $n_c$ or odd $n_a+n_b+2n_c$ and the ones
that make none of
$e^{b-a}$, $e^{c-a}$ and $e^{-a}$ divergent. Consequently, we have

\noindent
{\it{Type II}}: $E=0$ but $F\neq 0$.

The only solutions in this case are 
$(n_a, n_b, n_c)$ = (2, 2, 0) or (2, 2, 2) with either $\sin\alpha=0$ or 
$\cos\alpha=0$. The latter implies a horizon parallel  
to the plane $y=0$ and it may exist in the coordinates straight with respect 
to the $AdS_5$ horizon for a physical black hole. 
 
\noindent
{\it{Type III}}: $E=0$ and $F=0$.

The qualified solutions for $(n_a, n_b, n_c)$ are (2, 0, 0) and (2, 0, 2). 
They are also consistent with the other
components of the Einstein equations.
We have not found yet any restrictions on the shapes of the corresponding 
horizons.
The first solution of the integer triplet corresponds to the 
isotropic coordinates of a black hole
$(n_a=2, n_b=n_c=0)$, which can be obtained through a $y 
-independent$ coordinate transformation from the standard metric.  
It can be shown, however, that a $y-dependent$ and 
Gauss-normal form preserving transformation that leaves the brane 
intact does not exist. Therefore, this particular
horizon cannot be transformed into a pancake shaped one. 

For a general Gauss-normal form of the metric (1), we may define the 
$4\times 4$ matrix of ${\bar g}_{\mu\nu}$ by ${\cal G}$,
and the $R_{yy}-4\kappa^2$ equation can be
written as
$$
R_{yy}-4\kappa^2={1\over 2}\Big({{\partial}\over {\partial y}}
e^{-2\kappa y}
{{\partial}\over {\partial y}}\ln(-{\rm{det}}{\cal G})\Big)
+{1\over 4}\hbox{tr}{\cal G}^{-1}{{\partial{\cal G}}\over {\partial y}}
{\cal G}^{-1}{{\partial{\cal G}}\over {\partial y}}=0.
\numbereq\name{\eqavios}
$$
Now our statements regarding the Schwarzschild horizon can be generalized
to a spinning black
hole located on the brane $y=0$. Assume that
the 4-D Kerr metric \ref{\ker}{R. Kerr, \prl11 (1963) 237.} 
is recovered on 
the brane for sufficiently large $\kappa$, then we have 
$$
ds^2\simeq-{\Delta\over\rho^2}(dt-j\sin^2\theta d\phi)^2
+{\sin^2\theta\over\rho^2}[(r^2+j^2)d\phi-jdt]^2+{\rho^2\over\Delta}dr^2
+\rho^2d\theta^2
\numbereq\name{\eqdexios}
$$ 
for $y=0$,
where $\Delta=r^2-2GMr+j^2$ and $\rho^2=r^2+j^2\cos^2\theta$
with $M$ the mass 
and $j$ the angular momentum per unit mass. A horizon corresponds to 
the solutions of $\Delta=0$, that might
have two solutions or none. It is straightforward to 
show that the metric determinant
${\rm{det}}{\cal G}=-(r^2+j^2\cos^2\theta)^2
\sin^2\theta$ and its logarithm are free from the horizon singularities.
On the
other hand, the matrix ${\cal G}^{-1}{\partial{\cal G}\over\partial r}$ 
contains a ${1\over\Delta}$ singularity at the horizon.
If the 5-D extension 
of this horizon were bent towards or away from
the $y$ axis, this singularity 
would be shared by the matrix
${\cal G}^{-1}{\partial{\cal G}\over\partial y}$ 
off the brane. This is again forbidden by the Einstein equation 
(\eqavios). 

Having a pancake shaped horizon as the 5-D extension
of the Schwarszchild horizon
is physically implausible. If this were the case, somewhere
off the brane and 
near the horizon, we would expect that
$$
e^a\sim y-y_c, \qquad e^b\sim {1\over y-y_c},
\numbereq\name{\eqkentrwos}
$$
with $y_c$ a function of $r$. It follows then from equation
(\eqaristeros) that it would
take a finite amount of time for a light beam coming out of the horizon 
to propagate in the $y$-direction.
Consequently the black hole would appear to be
luminating in the $y$-direction.

The conclusions we have reached above depend
on the form of the metric, equation (\eqena), and the Einstein equations.
We might try to relax the Gauss normal form of the metric. In the case 
of a spherical black hole, we might consider the metric
$$
ds^2=e^{-2\kappa|y|}(-e^udt^2+e^vdr^2+r^2d\Omega^2)+e^fdy^2,
\numbereq\name{\eqvasilos}
$$
where $u, v, f$ are functions of $r$ and $y$.
The equation $R_{yy}-4\kappa^2=0$ then becomes
$$
\ddot u+\ddot v+{1\over 2}(\dot u^2+\dot v^2)+e^{2\kappa y+f-v}
(f^{\prime\prime}+{1\over 2}f^{\prime 2})=0.
\numbereq\name{\eqpaok}
$$
Let's assume that an approximate Schwarszchild metric can
be recovered on the brane, which is located at
$y=0$ and that the 5-D
extension of the horizon is given by $H(r,y)=0$. If we
consider an arbitrary point 
on it, $P(r_0,y_0)$, with the variable $\xi$ defined as before, we have 
$$
u\sim \ln\xi \qquad v\sim -\ln\xi \qquad f\sim n_c\ln\xi
\numbereq\name{\eqkati}
$$
with $n_c$ being an even integer. Then,
$$
\ddot u+\ddot v+{1\over 2}(\dot u^2+\dot v^2)
\sim {1\over \xi^2}\sin^2\alpha
\numbereq\name{\eqalvez}
$$
and
$$
e^{2\kappa y+f-v}(f^{\prime\prime}+{1\over 2}f^{\prime 2})
\sim (-n_c+{1\over 2}n_c^2)\xi^{n_c+1-2}\cos^2\alpha.
\numbereq\name{\eqagios}
$$
For $n_c<-1$, (\eqagios) represents the leading
singularity of the equation (\eqpaok), the only 
solution is $\cos\alpha=0$ and this horizon will not join the Schwarszchild 
horizon on the brane. If $n_c>-1$, (\eqalvez) represents
the leading singularity, 
and the only solution is $\sin\alpha=0$. Therefore, the 5-D extension of 
Schwarszchild horizon is again a tube.

The existence of a 5-D extension of a 4-D
Schwarzschild horizon for a 
physical black hole relies on the recovery
of 4-D general relativity on the 
brane for large $\kappa$. The rigorous statements we made on the possible 
shapes of the horizon, however, are independent of
the value of $\kappa$, since the terms
of the first derivative with respect to $y$ in
equations (\eqevon)-(\eqefta) do not contribute to the 
leading singularity of a horizon. In the following, we illustrate various 
types of horizons for a 5-D Schwarzschild metric $(\kappa=0)$, i. e.,
$$
ds^2=-\Big(1-{l^2\over R^2}\Big)dt^2+{dR^2\over 1-{l^2\over R^2}}+R^2
(d\chi^2+\sin^2\chi d\Omega^2),
\numbereq\name{\eqesia}
$$
where $l$ is the Schwarzschild radius, $R$ is the radial coordinate,
and $\chi$ is a polar angle on $S^3$, $d\Omega$ is the solid angle 
on $S^2$. If we introduce cylindrical coordinates
$r^\prime=R\sin\chi$ and $y^\prime=R\cos\chi$, we find
$$
ds^2=-\Big(1-{l^2\over R^2}\Big)dt^2+{l^2\over 1-{l^2\over R^2}}
{(r^\prime dr^\prime+y^\prime dy^\prime)^2\over R^2}+dr^{\prime2}
+r^{\prime2}d\Omega^2+dy^{\prime2},
\numbereq\name{\eqnolb}
$$
and we have a circular horizon on $r^\prime-y^\prime$ plane
at the expense of introducing 
off-diagonal terms of the metric. To enforce the Gauss-normal form of the 
metric, consider the transformation 
$$
R=R(r,y) \qquad \chi=\chi(r,y).
\numbereq\name{\eqnocxi}
$$
The functions $R(r,y)$ and 
$\chi(r,y)$ have to satisfy the conditions
$$
\eqalign{
{R^2\over R^2-l^2}\Big({\partial R\over\partial y}\Big)^2
+R^2\Big({\partial\chi\over\partial y}\Big)^2&=1\cr
{1\over R^2-l^2}{\partial R\over\partial r}{\partial R\over\partial y}
+{\partial\chi\over\partial r}{\partial\chi\over\partial y}&=0.\cr}
\numbereq\name{\eqnopoiu}
$$
There is a simple solution with $R={\sqrt{y^2+l^2}}$
and $\chi$ an arbitrary 
function of $r$ only, which results in a Gauss-normal form of the
metric (\eqnolb), i. e.,
$$
ds^2=-{y^2\over y^2+l^2}dt^2+(y^2+l^2)\Big({d\chi\over dr}\Big)^2dr^2
+(y^2+l^2)\sin^2\chi d\Omega^2+dy^2.
\numbereq\name{\eqnoash}
$$
The horizon, $y=0$, then becomes of type III. This transformation,
however, is singular 
on the horizon, more specifically the jacobian
${{\partial}(R, \chi)\over {{\partial}(r, y)}}$
is zero. If a solution to equations
(\eqnopoiu) that is nonsingular at the horizon could be found,
then we would have a horizon 
of type I, since $\lim_{R\to l}{\partial R\over\partial y}=0$ in any 
case in order to balance the first equation of (\eqnopoiu) and the 
nonvanishing jacobian demands that 
$\lim_{R\to l}{\partial R\over\partial r}\neq 0$. 

In summary, we have found that the vacuum Einstein equations together with 
a Gauss-normal form of the metric place fairly stringent restrictions on 
the form of the event horizon.
The suggested pancake shaped extension of the 4-D Schwarzschild horizon
cannot exist in the Gauss-normal form of the metric. In the
coordinate system based on the brane \ref{\gar}{J. Garriga and
T. Tanaka, \prl84 (2000) 2778.},
the event horizon
has a tubular shape extending possibly to infinity in
agreement with recent numerical studies \ref{\shir}
{T. Shiromizu and M. Shibata, {\it Black holes in the
brane world: Time symmetric initial data}, hep-th/0007203;
A. Chamblin, H. Reall, H. Shinkai
and T. Shiromizu, {\it Charged Brane-World Black Holes},
hep-th/0008177, N. Dadhich, R. Maartens, P. Papadopoulos
and V. Rezania, \pl487 (2000) 1.}, while
in the coordinate system based on the $AdS_5$
horizon, the region of validity of linear gravity excludes
this possibility. The horizon then might belong to
either Type II or Type III. 
The technique of balancing the leading singularities in
the Einstein equation that
we have developed in the present
work may be generalized to other singularities
of the solution, eg. the curvature singularity. Since
our analysis does not depend on the value of $\kappa$,
our classification of horizons applies to the large
extra compact dimension scenario $M^4 \times S^1$,
which corresponds to $\kappa=0$ and periodicity in the
$y$-direction. Our rigorous statements
on the shape of the horizon provide not only hints to the
form of the exact solution, but may also impact on the stability
of it. 
It is also interesting to explore 
the metric of a physical black hole numerically. We hope to report our 
progress towards these directions in the near future.

\vskip .1in
\noindent
{\bf Acknowledgments.} \vskip .01in \noindent
We would like to thank N. Khuri, J. T. Liu, B. Morariu, A. Polychronakos,
M. Porrati, C. Pope, T. Tomaras and Y. S. Wu
for useful discussions.
This work was supported in part by the Department of Energy under Contract
Number DE-FG02-91ER40651-TASK B.

\noindent
\immediate\closeout1
\bigbreak\bigskip

\line{\twelvebf References. \hfil}
\nobreak\medskip\vskip\parskip

\input refs

\vskip 1in

\input epsf
\epsfxsize 3truein
\centerline{\epsffile{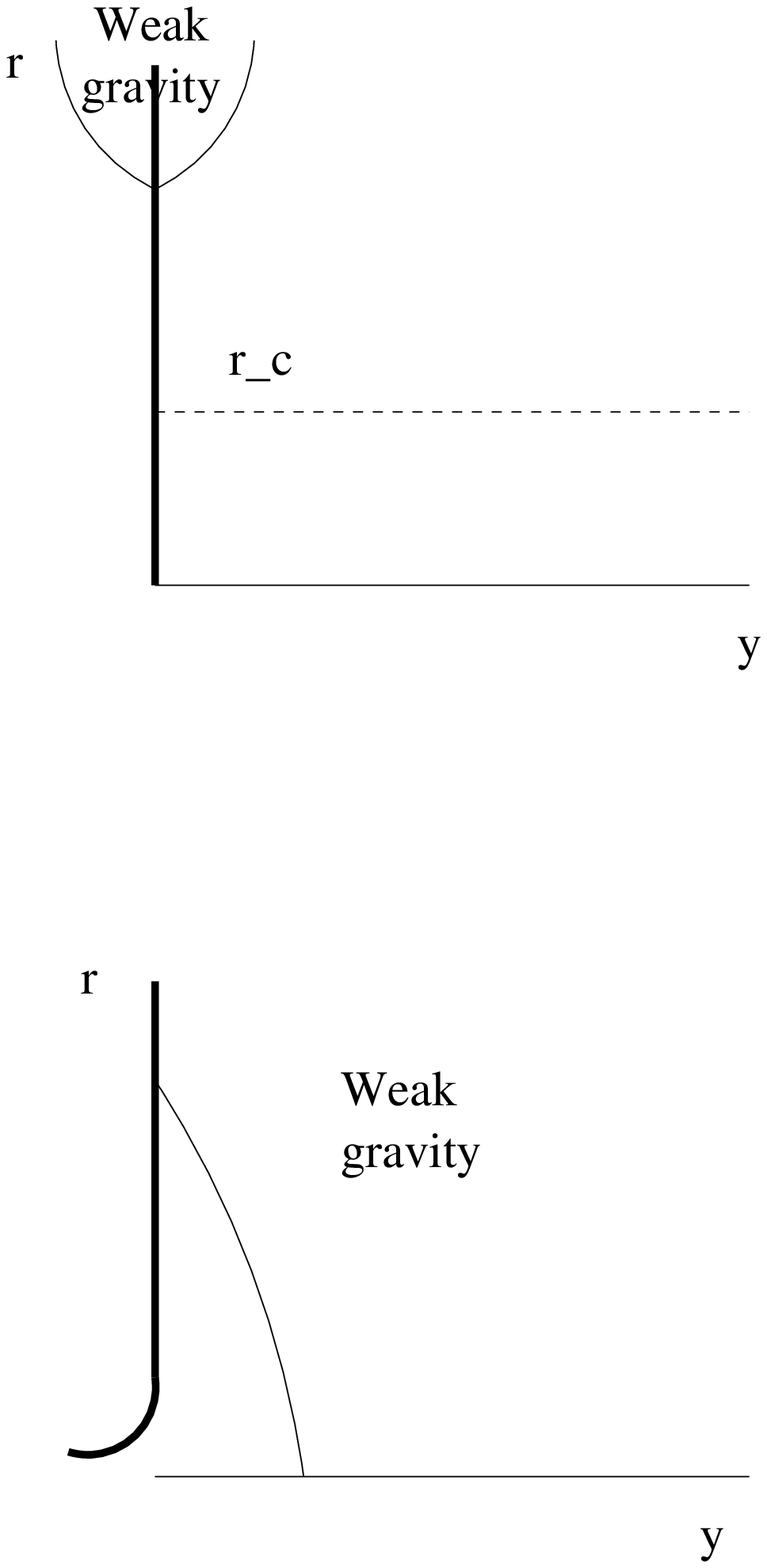}}
\medskip
\noindent
Figure 1
\medskip
A physical black hole viewed in two coordinate systems:
\smallskip
a) In the coordinates based on the brane, where the thick line
denotes the brane and the dashed line the off-brane extension
of the 4-D Schwarzschild horizon. The Schwarzschild radius is
$r_c=2GM(1+O({1\over {\kappa^2G^2M^2}}))$ and the mass point
is located at $r=y=0$.
\smallskip
b) In the coordinates based on the $AdS_5$ horizon, where
the thick line denotes the bent brane.

\vfil\end

\bye